\documentclass[twocolumn,showkeys,preprintnumbers,amsmath,amssymb,epsfig,floatfix,prl]{revtex4-1}
\usepackage{amsmath}
\usepackage{mathtools}
\usepackage{physics}
\usepackage[version=4]{mhchem}
\usepackage[english]{babel}
\usepackage[utf8]{inputenc}
\DeclareMathAlphabet{\altmathcal}{OMS}{cmsy}{m}{n}
\usepackage{siunitx}
\usepackage{multirow}
\usepackage{amsfonts}
\usepackage{amssymb}
\usepackage{calligra}
\usepackage{calrsfs}
\DeclareMathAlphabet{\mathcalligra}{T1}{calligra}{l}{m}
\usepackage{epsfig}
\usepackage{graphicx}
\usepackage{dcolumn}
\usepackage{color}
\usepackage{natbib}  
\usepackage{hyperref}
\usepackage{breakurl}
\hypersetup{colorlinks=true, citecolor=blue, urlcolor=blue, linkcolor=blue}
\usepackage{bm}
\usepackage{booktabs}
\newcolumntype{C}{>{$}c<{$}}
\AtBeginDocument{
\heavyrulewidth=.08em
\lightrulewidth=.05em
\cmidrulewidth=.03em
\belowrulesep=.65ex
\belowbottomsep=0pt
\aboverulesep=.4ex
\abovetopsep=0pt
\cmidrulesep=\doublerulesep
\cmidrulekern=.5em
\defaultaddspace=.5em
}

\newcolumntype{L}[1]{>{\raggedright\arraybackslash}p{#1}}
\newcolumntype{C}[1]{>{\centering\arraybackslash}p{#1}}
\newcolumntype{R}[1]{>{\raggedleft\arraybackslash}p{#1}}

\begin{document}
\title{Enhanced Magnetism and Phase Transitions in Ultrathin Quantum Spin Liquid \ce{Na2IrO3} Flakes}
\author{Deepak K. Roy}
\author{Mukul Kabir}
\email{mukul.kabir@iiserpune.ac.in}
\affiliation{Department of Physics, Indian Institute of Science Education and Research, Pune 411008, India}
\date{\today}

\begin{abstract}
{\bf The quest for quantum spin liquids has garnered significant attention due to their rich physics and disruptive prospects in quantum communication and computation. Spin-orbit coupling, electron correlation, and structural distortion play critical roles in the candidate materials that eventually order antiferromagnetically at low temperatures. We introduce quantum electron confinement to the existing complexity and explore the interplay between Heisenberg and Kitaev interactions in ultrathin \ce{Na2IrO3} layers using first-principles calculations. The zigzag antiferromagnetic state in the monolayer is reinforced and pushed further away from the Kitaev spin liquid state due to the increased strength of Heisenberg and off-diagonal exchange interactions. In contrast, the carrier-doped flakes undergo a Mott insulator-to-metal transition accompanied by an antiferromagnetic to ferromagnetic transition. These findings present exciting prospects for comprehending magnetism in a novel two-dimensional framework of non-van der Waals correlated oxide flakes.} 
\end{abstract}
\maketitle

Quantum spin liquid (QSL) is an exotic state of matter where highly correlated spins evade symmetry-breaking phase transition down to the lowest temperatures.~\citep{nature08917,annurev-conmatphys-031218-013401,s42254-019-0038-2}  QSLs exhibit long-range entanglement, topological order, emergent gauge fields and fractional excitations that attracted enormous attention in quantum condensed matter physics.   
In 1973, Anderson proposed a two-dimensional QSL state in a triangular $S=1/2$ Heisenberg spin system interacting antiferromagnetically.~\citep{ANDERSON1973153} The conjectured resonating valence bond ground state consists of a quantum superposition of spin singlets, and the elementary excitations are the fractionalized $S=1/2$ spinons that can propagate in the lattice. However, the detection of these fractionalized excitations in real materials remains inconclusive despite extensive efforts.~\citep{PhysRevLett.91.107001,PhysRevB.77.104413,PhysRevLett.100.087202,science.1188200,nature11659,science.aab2120}
In contrast, the Kitaev QSL state is an exact solution to a model Hamiltonian with bond-specific $S=1/2$ nearest-neighbour Ising interactions, $K_{\gamma}S_i^{\gamma}S_j^{\gamma}$, where $\gamma$ denotes different bonds.~\citep{KITAEV20062} In this model, the spin fractionalizes into itinerant and localized Majorana fermions. 

Realizing Kitaev QSL state in real materials is challenging due to the existence of additional spin interactions, such as the Heisenberg exchange.~\citep{PTPS.160.155,PhysRevLett.102.017205,PhysRevLett.105.027204,PhysRevLett.110.097204,PhysRevLett.112.077204} In this context, spin-orbit entangled pseudospins on a honeycomb lattice can generate Kitaev interaction in the edge-shared octahedral Mott insulators. The interplay between correlation and spin-orbit coupling makes \ce{Ir^4+} oxides and \ce{Ru^3+} chlorides the primary candidate materials for hosting the Kitaev QSL state.~\citep{PhysRevB.82.064412, PhysRevLett.108.127203, PhysRevB.93.134423, PhysRevB.92.235119,nature25482}
However, magnetic ordering is observed at low temperatures in two-dimensional honeycomb iridates and \ce{RuCl3},~\citep{PhysRevB.82.064412, PhysRevLett.108.127203,PhysRevB.93.134423, PhysRevB.92.235119} while the existence of the QSL state in \ce{H3LiIrO6} remains a topic of debate.~\citep{nature25482} Nevertheless, the ordering temperature is an order of magnitude lower than the energy scale of magnetic interactions, indicating the presence of magnetic frustration. 

Among the iridates, \ce{Na2IrO3} has received considerable attention since it was envisioned for realizing Kitaev physics.~\citep{PTPS.160.155,PhysRevLett.102.017205} Due to the extended \ce{Ir}$-5d$ orbitals, iridates reside far away from the Mott insulating limit, $U \gg W$, with $U$ and $W$ being the electron correlation and electronic bandwidth, respectively. However, owing to strong spin-orbit coupling $\lambda_{\ce{Ir}} \sim$ 400 meV, \ce{Na2IrO3} is a weak spin-orbital entangled Mott insulator similar to \ce{Sr2IrO3}.~\citep{PhysRevLett.109.266406,PhysRevLett.117.187201,PhysRevB.98.125117,PhysRevLett.101.076402,science.1167106} The half-filled $j_{\rm eff}=1/2$ narrow band splits due to electron correlation, and the singly-occupied Kramers doublet describes the low-energy physics. Consequently, the delicate interplay between comparable $U, W,$ and $\lambda$ becomes intriguing. 

Given the spin-orbital nature, the interactions between $j_{\rm eff}=1/2$ pseudospins are highly anisotropic,~\citep{PTPS.160.155,PhysRevLett.102.017205} 
and the conventional Heisenberg exchange is suppressed according to the Jackeli-Khaliullin (JK) mechanism~\citep{PTPS.160.155,PhysRevLett.102.017205}. The hopping between the adjacent $j_{\rm eff}=1/2$ and virtual $j_{\rm eff}=3/2$ ($m_j = \pm 3/2$) orbitals dominates, leading to anisotropic Ising ferromagnetic interaction due to Hund's coupling. The three bonds emerging from each metal site accommodate Kitaev interaction for edge-shared octahedra. The easy-axes of the $j_{\rm eff}=1/2$ pseudospin are perpendicular to \ce{Ir-O2-Ir} bond planes, and thus the Ising axes are orthogonal to each other for the three bonds.
However, trigonal distortions of the cubic octahedral environment lift the $t_{2g}$ orbital degeneracy and partially quench the orbital angular momentum, leading to a decrease in the Kitaev interaction and an increase in Heisenberg exchange.~\citep{PhysRevLett.105.027204,Winter_2017} Accordingly, at temperatures below 15 K, the bulk \ce{Na2IrO3} exhibits antiferromagnetic (AFM) ordering, forming a zigzag spin structure that is consistent with a large Kitaev exchange.~\citep{PhysRevB.82.064412,PhysRevB.85.180403,PhysRevB.83.220403} Thus, a Heisenberg-Kitaev model is deemed appropriate.~\citep{PhysRevLett.102.017205,PhysRevLett.105.027204} 

We explore magnetism in ultrathin layers of \ce{Na2IrO3} within the first-principles calculations. Such ultrathin flakes have recently been synthesised through chemical exfoliation, presenting a promising avenue for investigating frustrated magnetism in two-dimensional systems with strong correlation and spin-orbit coupling. Surprisingly, we find that magnetism not only survives but is reinforced up to the monolayer limit. Similar to the bulk counterpart, the ultrathin and passivated \ce{Na2IrO3} flakes exhibit a Mott-insulating zigzag AFM ground state. The present investigation focuses on the evolution of different magnetic interactions from the bulk crystal to the monolayer. Further, chemically exfoliated flakes may be charge-doped, and our findings suggest that it drives an intriguing magnetic phase transition along with a Mott insulator-to-metal transition. These results present an exciting opportunity for further exploration.

\subsection{Relativistic Mott insulating bulk}
\begin{figure}[!t]
\begin{center}
{\includegraphics[width=0.48\textwidth, angle=0]{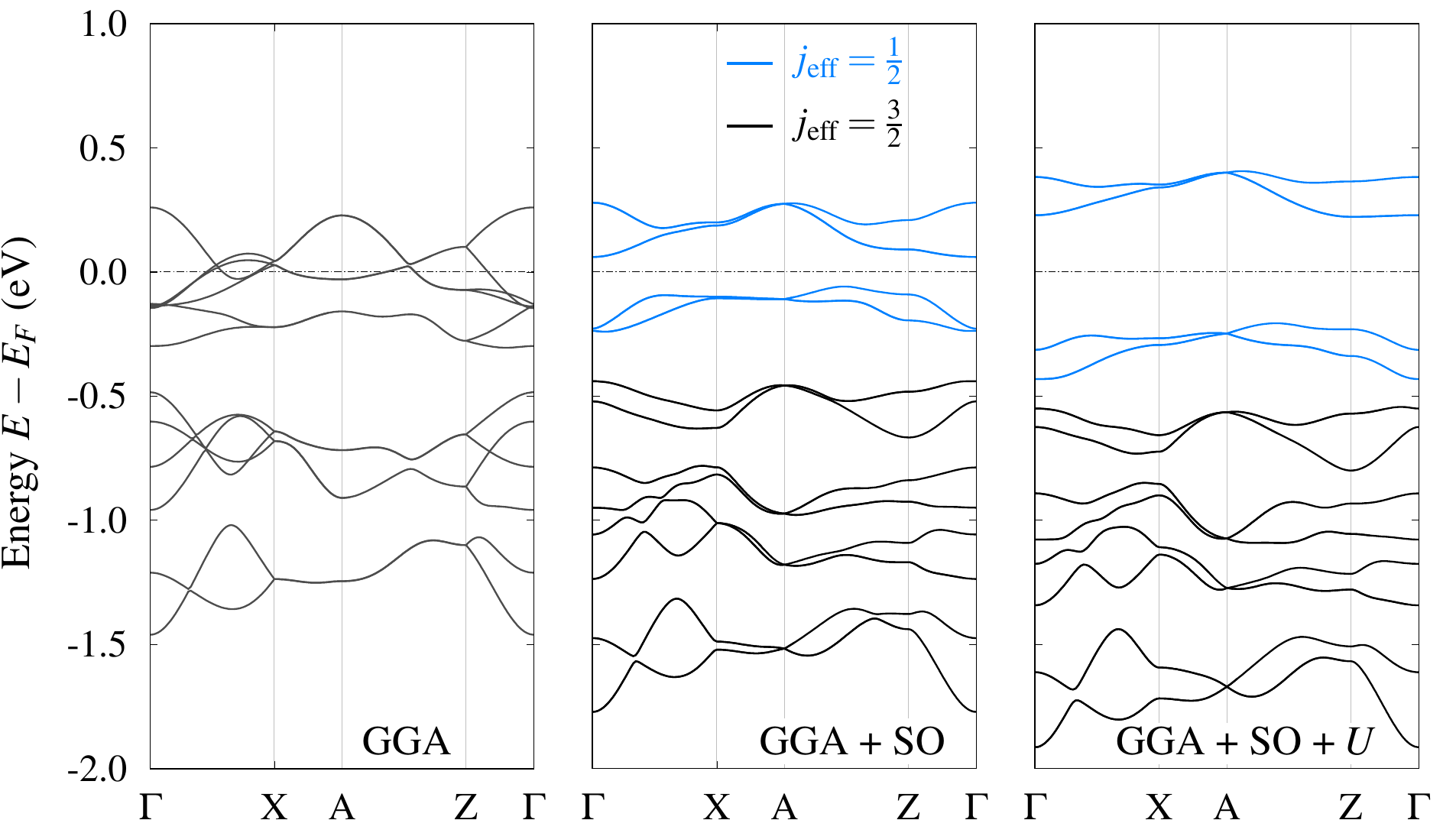}}
\caption{\ce{Ir} $5d-t_{2g}$ band structure obtained with GGA, GGA + SO and GGA + SO + $U_{\ce{Ir}}$ ($U_{\ce{Ir}}$ = 1 eV, $J_H$ = 0.5 eV). Structural distortion and SO coupling play critical roles. \ce{Na2IrO3} is a relativistic Mott insulator since SO coupling opens up an insulating gap between the occupied and unoccupied $j_{\rm eff} = 1/2$ bands that further increases with $U_{\ce{Ir}}$. The calculated gap agrees well with experimental data.~\citep{PhysRevLett.109.266406,PhysRevLett.110.076402,PhysRevB.88.085125} 
}
\label{fig:figure1}
\end{center}
\end{figure}

Bulk \ce{Na2IrO3} crystallizes in the monoclinic $C2/m$ phase, and the calculated lattice parameters ($a=\SI{5.45}{\angstrom}$, $b=\SI{9.43}{\angstrom}$, $c=\SI{5.64}{\angstrom}$) agree with the experimental data.~\citep{PhysRevLett.108.127204} The crystal structure comprises two-dimensional layers of edge-shared \ce{IrO6} octahedra that construct a honeycomb \ce{Ir}-lattice in the $ab$-plane, with Na atoms at the centre. These layers are stacked along the $c$-axis by alternating with pure Na-layers. Since the $5d$ orbitals are more extended than the $3d$ orbitals, the resonant inelastic X-ray scattering (RIXS) data corroborates a much larger octahedral crystal field splitting $\Delta_{\rm OC}$ about 3 eV in Ir-compounds.~\citep{PhysRevLett.110.076402} Thus, with $\Delta_{\rm OC} \sim$ 3.4 eV in \ce{Na2IrO3}, the $5d^5$ electrons of \ce{Ir^4+} ions exist in the low-spin $t_{2g}^5$ configuration, and the state can alternatively be described with one $t_{2g}$ hole. The dispersion of the \ce{Ir} $5d$-$t_{2g}$ manifold is broad, around 2 eV, and is impacted by structural distortion. This leads to narrower $t_{2g}$ sub-bands, with $W$ in the range of 0.5$-$0.6 eV (Figure~\ref{fig:figure1}). A metallic solution characterized by a high density of states at the Fermi level is obtained using GGA, and it remains metallic even with the inclusion of the Hubbard correlation $U_{\ce{Ir}}$ term in the calculation. An insulating gap appears only when the spin-orbit coupling is introduced (Figure~\ref{fig:figure1}). With the strong spin-orbit coupling, the degeneracy of the threefold \ce{Ir}-$t^5_{2g}$ orbitals is further lifted, resulting in states with effective angular momentum $j_{\rm eff}=3/2$ and $j_{\rm eff}=1/2$. While the $j_{\rm eff}=3/2$ manifold is completely filled, the half-filled $j_{\rm eff}=1/2$ splits into lower and upper Hubbard bands due to electron correlation. Thus, \ce{Na2IrO3} is a relativistic Mott insulator (Figure~\ref{fig:figure1}). The insulating gap increases from 120 meV in GGA + SO to 430 meV as an additional on-site Coulomb interaction, $U_{\rm Ir}=$ 1 eV, is incorporated into the calculation (Figure~\ref{fig:figure1}). The results are consistent with the angle-resolved photoemission spectroscopy, optical conductivity, and RIXS data,~\citep{PhysRevLett.109.266406,PhysRevLett.110.076402,PhysRevB.88.085125} and previous theoretical estimations.~\citep{PhysRevLett.109.266406,PhysRevB.88.085125,PhysRevB.83.220403,PhysRevB.93.195135,PhysRevB.91.161101}  The effective correlation, which is determined from the occupied and unoccupied $j_{\rm eff}=1/2$ states, ranges from 0.3 to 0.6 eV for $U_{\ce{Ir}}$ values of 0 to 1 eV, consistent with the experimental estimation of 0.52 eV.~\citep{PhysRevB.98.125117} The bandwidths of the $t_{2g}$ subbands are further affected by the SO coupling and become narrower $\sim$ 200 meV near the Fermi level, compatible with the ARPES experiment.~\citep{PhysRevLett.109.266406} Therefore, we conclude that \ce{Na2IrO3} resides at the borderline of Mott criterion, $U > W$.  

Structural distortions play a central role in determining the quantum state. The cubic symmetry of the \ce{IrO6} octahedra is broken as trigonal and orthorhombic distortions are inherited,~\citep{PhysRevLett.108.127204,PhysRevB.85.180403} and the \ce{Ir-O-Ir} angles $\theta_{\ce{Ir}}$ become more than $\ang{90}$. The computed average $\theta_{\ce{Ir}} \sim  \ang{100.3}$ agrees with the measured range $98-\ang{100}$.~\citep{PhysRevLett.108.127204} The crystal-filed splitting $\Delta_t$ of the $j_{\rm eff}=3/2$ states, induced by the trigonal distortion, is determined to be 110 meV in high-resolution RIXS measurements.~\citep{PhysRevLett.110.076402} The present result of $\Delta_t \sim$ 170 meV corroborates the experimental data, and since $\Delta_t < \lambda_{\ce{Ir}}$, the microscopic model with $j_{\rm eff}$ picture remains valid. Further, the spin-orbit excitation ($j_{\rm eff}$-${3/2}$ $\rightarrow$ $j_{\rm eff}$-${1/2}$) energy of 0.7 eV within GGA + SO + $U_{\ce{Ir}}$ compares well with the 
RIXS peak located at 0.7-0.8 eV.~\citep{PhysRevLett.110.076402}

\begin{figure*}[!t]
\begin{center}
{\includegraphics[width=0.98\textwidth, angle=0]{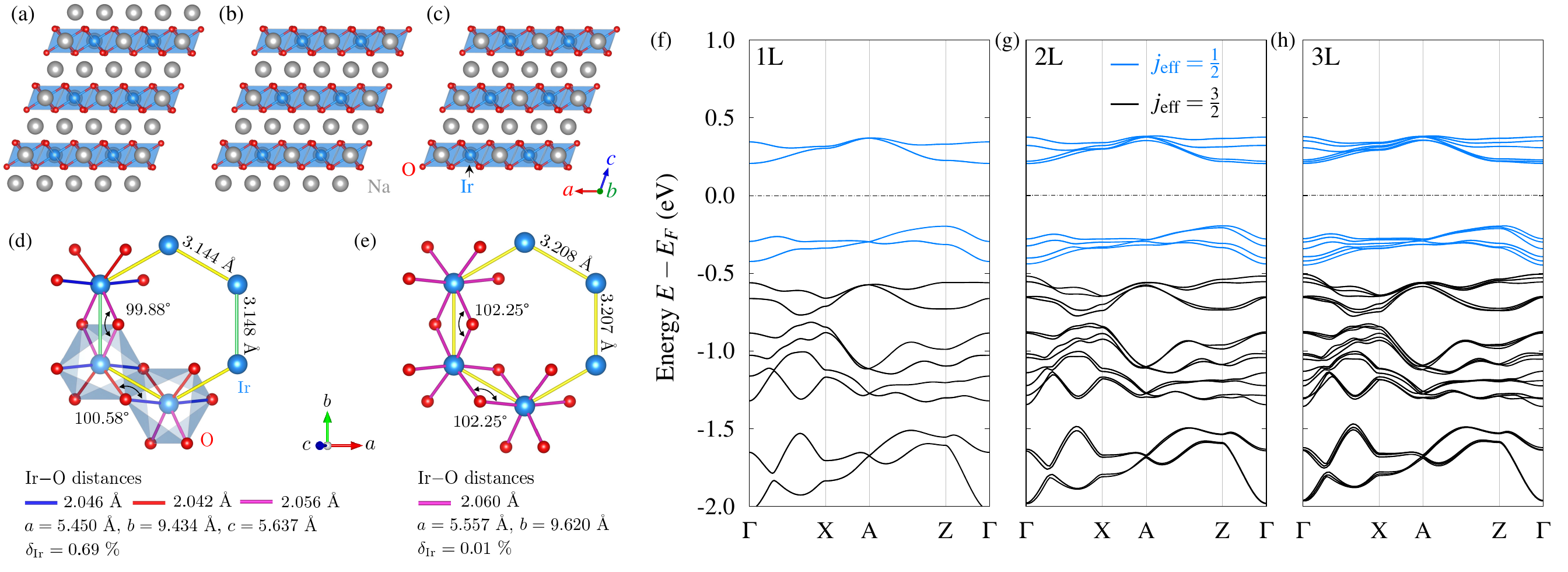}}
\caption{Crystal structure details and electronic band structure of ultrathin flakes. (a)-(c)
Various surface terminations are possible for the ultrathin flakes, shown for three-layer (3L) \ce{Na2IrO3}. The bulk structure consists of alternating stacking of honeycomb \ce{NaIr2O6} layers separated by hexagonal \ce{Na3} layers, and the flakes with and without the pure \ce{Na3} termination are equally probable. Flake where (a) both sides [\ce{Na3}$|$$\cdots$$|$\ce{Na3}], (b) one side  [$\otimes$$|$$\cdots$$|$\ce{Na3}] and (c) none of the sides [$\otimes$$|$$\cdots$$|$$\otimes$] are terminated with Na-layer. We also explore the chemically passivated flakes, which are not shown in the figure. Similar structures for the monolayer and bilayers are studied. The local structural details of the honeycomb \ce{NaIr2O6} layer in (d) bulk and (e) passivated monolayer, [\ce{H-Na3}$|$\ce{NaIr2O6}$|$\ce{Na3-H}]. For clarity, the Na-atoms are not shown. Same color bonds refer to the equivalent \ce{Ir-O} distances. As the thickness decreases from the bulk to passivated 1L, the in-plane lattice parameters increase, and \ce{Ir-O} bonds, \ce{Ir-O-Ir} angles, and \ce{Ir-Ir} distances become uniform. Therefore, the Ir-off-centering distortion $\delta_{\rm Ir}$ becomes negligibly small in the monolayer. (f)-(h) Electronic band structure of passivated flakes, which remain spin-orbit coupled Mott insulator. The electronic structure parameters, such as band gaps, bandwidths and electron correlation, remain equivalent to those for the bulk \ce{Na2IrO3}.
}
\label{fig:figure2}
\end{center}
\end{figure*}

According to the JK mechanism, the magnetic superexchange interactions between \ce{Ir^4+} ions with $j_{\rm eff}=1/2$ pseudospins are essentially of Kitaev type for edge-shared, undistorted octahedra.~\citep{PTPS.160.155,PhysRevLett.102.017205} Trigonal distortions perturb the JK mechanism and lift the $t_{2g}$ degeneracy, resulting in partial quenching of the orbital moment. Consequently, bulk \ce{Na2IrO3} exhibits magnetic ordering below 15 K~\citep{PhysRevB.82.064412,PhysRevB.85.180403,PhysRevB.83.220403} as the channels for Heisenberg and off-diagonal interactions are introduced. The inelastic neutron scattering,~\citep{PhysRevLett.108.127204} X-ray diffraction,~\citep{PhysRevB.85.180403} and RIXS experiments~\citep{PhysRevB.83.220403} indicate an AFM zigzag ordering. In concurrence, we find that the zigzag AFM ground state is 4 meV/\ce{Ir} lower than the ferromagnetic (FM) case, while the stripe and N\'eel AFM states remain 5 and 6 meV/\ce{Ir} higher in energy. The spin moment of 0.28 $\mu_{\rm B}$/\ce{Ir} is significantly smaller than expected for the $j_{\rm eff}=1/2$ system, in consensus with the experimental result.~\cite{PhysRevB.85.180403} Such reduction in the spin moment is also observed in other \ce{Ir^4+} compounds, \ce{Sr2IrO4}, \ce{Ba2IrO3}, and $\beta$-\ce{Li2IrO3}.~\citep{PhysRevB.84.100402,PhysRevLett.105.216407,s41467-017-01071-9} The reduction in the spin moment is attributed to the $5d$-$2p$ hybridization and small exchange splitting in the $5d$ bands. While the orbital moment $\mu_{\rm o} \sim$ 0.34 $\mu_{\rm B}$/\ce{Ir} is equivalent across the magnetic states, the spin moment is highly dependent, $\mu_{\rm s} \sim$ 0.1, 0.2, and 0.4 $\mu_{\rm B}$/\ce{Ir} for  N\'eel, stripe, and FM arrangements. Further, $\mu_{\rm s}$ and $\mu_{\rm o}$ moments are weakly noncollinear and lie in the $ac$-plane at an angle with the crystallographic $a$ axis, which is in qualitative agreement with the experimental~\citep{nphys3322} and earlier theoretical results.~\citep{PhysRevLett.115.167204}

\subsection{Ultrathin \ce{Na2IrO3} flakes}
Exploring ultrathin layers of \ce{Na2IrO3} could provide fascinating insights, as they offer additional complexity due to the quantum electron confinement, in addition to the existing electron correlation and spin-orbit coupling. These layers have recently been chemically exfoliated, and a comprehensive theoretical exploration of the same  is highly relevant. Through chemical exfoliation, the weaker \ce{Na-O} bonds between the \ce{Na3} and \ce{NaIr2O6} layers are cleaved. As a result, the flakes with and without the pure Na-layer terminations are equally likely,~\citep{PhysRevB.91.041405,PhysRevB.96.161116} and we study all such possibilities, some of which are shown in Figure~\ref{fig:figure2}(a)-(c). Ultrathin layers are cleaved parallel to the honeycomb \ce{IrO} layers resulting in two-dimensional flakes in the crystallographic $ab$-plane. The calculated phonon dispersions indicate that the neutral and charge-doped ultrathin non-vdW flakes are dynamically stable. Although a flexural phonon mode with small imaginary frequencies can be observed in a small pocket near the Brillouin zone centre, it is understood to be linked to the difficulties in achieving numerical convergence in two-dimensional materials.~\citep{PhysRevB.89.205416,PhysRevMaterials.3.074002} Since exfoliated flakes are either \ce{Na}-rich or \ce{Na}-deficient, they are electron- or hole-doped at the surface and will have an intriguing effect on their physical properties. 

\begin{figure}[!t]
\begin{center}
{\includegraphics[width=0.45\textwidth, angle=0]{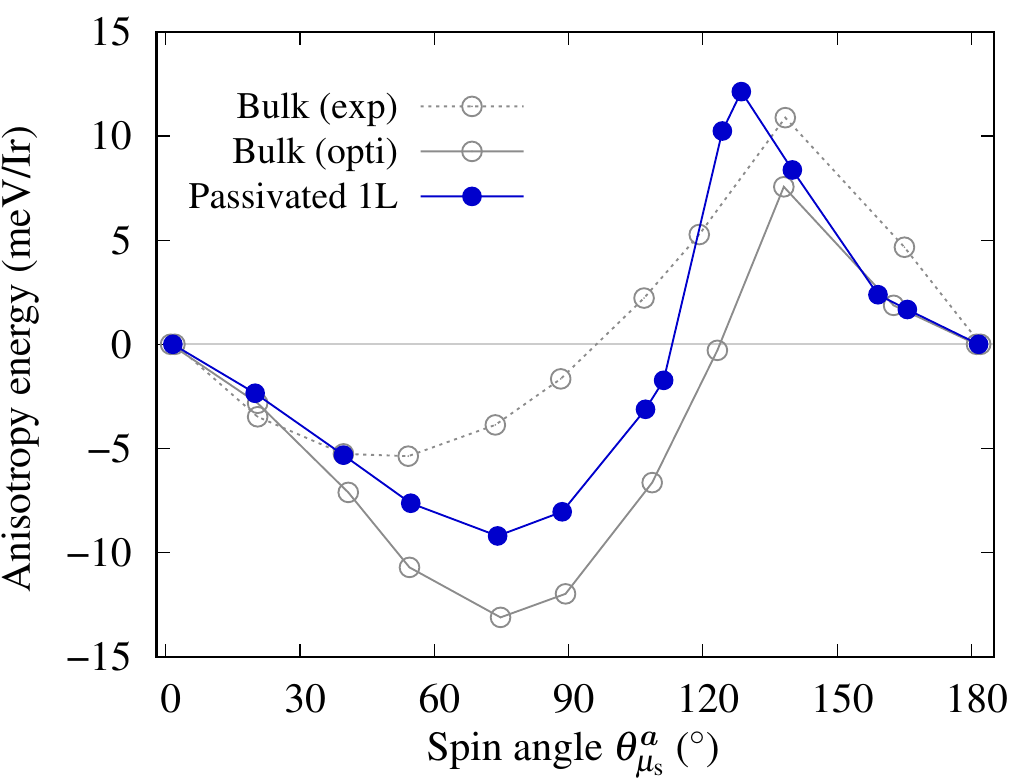}}
\caption{The magnetic anisotropy energy in the crystallographic $ac$ plane is illustrated as the spin moment rotates at an angle $\theta_{\mu_{\rm s}}^{\bm a}$ with respect to the $\bm{a}$-axis. The calculations are performed using GGA + SO + $U_{\rm Ir}$ + $J_{\rm H}$ approach in the zigzag AFM order, and the obtained energies are presented relative to the energy with the moment oriented along the $\bm{a}$-axis. The moments in the Mott insulating zigzag AFM ground state are aligned at an angle with the $\bm{a}$-axis.  
}
\label{fig:figure5}
\end{center}
\end{figure}

We begin by analyzing ultrathin layers that are terminated with \ce{Na3} layers on both sides, which are further H-passivated, creating a structure of [\ce{H-Na3}$|$$\cdots$$|$\ce{Na3-H}]. Bader charge analysis demonstrates that the magnetic honeycomb \ce{NaIr2O6} sublayers in these flakes have identical charge distribution as in bulk \ce{Na2IrO3} and are charge neutral. The in-plane lattice parameters increase consistently due to thickness reduction and are 2\% longer in the monolayer than in bulk [Figure~\ref{fig:figure2}(d)-(e)]. The lattice distortions are also affected as the Ir-off-centring distortion $\delta_{\rm Ir}$ substantially decreases from 0.69\% in bulk to 0.01\% in the passivated monolayer. In contrast, in the monolayer, trigonal distortion increases $\theta_{\rm Ir} \sim \ang{102.3}$. The overall electronic structure of the passivated flakes remains similar to that of the bulk and retains the relativistic Mott insulating state with about 400 meV band gaps [Figure~\ref{fig:figure2}(f)-(h)]. Calculated $\Delta_{\rm OC}$ = 3.4 eV, the effective correlation calculated from the upper and lower Hubbard bands of 0.6 eV and bandwidths near the Fermi level, 160$-$250 meV, also remain similar to bulk values. 

Surface passivated ultrathin flakes with one to three \ce{NaIr2O6} sublayers remain magnetic and exhibit a zigzag AFM configuration, similar to the bulk. These findings are surprising because they contradict the conventional wisdom that ordering decreases as the thickness decreases.~\citep{s41565-019-0438-6} This is particularly unexpected given that the bulk material orders only below 15 K.  Further, the zigzag ordered \ce{Ir}-moments exhibit a curious characteristic of not aligning with any of the crystallographic axes. In passivated monolayer and optimized bulk lattice, the $\ce{Ir}$ moments are situated in the $ac$-plane at an angle of approximately $\ang{75}$ and $\ang{70}$, respectively, with respect to the crystallographic $\bm{a}$-axis (Figure~\ref{fig:figure5}). In comparison, calculations reveal that these angles are somewhat smaller for \ce{Na2IrO3} with experimental lattice. Moreover, the $\mu_{\rm s}$ and $\mu_{\rm o}$ moments exhibit a weakly noncollinear behavior with mutual angles that are below $\ang{10}$. These observations agree with both experimental~\citep{nphys3322} and theoretical results.~\citep{PhysRevLett.115.167204} The passivated monolayer flake exhibits a similar trend as the bulk (Figure~\ref{fig:figure5}), although with a greater degree of noncollinearity between the spin and orbital moments, which is approximately $\ang{25}$.

\begin{figure}[!t]
\begin{center}
{\includegraphics[width=0.40\textwidth, angle=0]{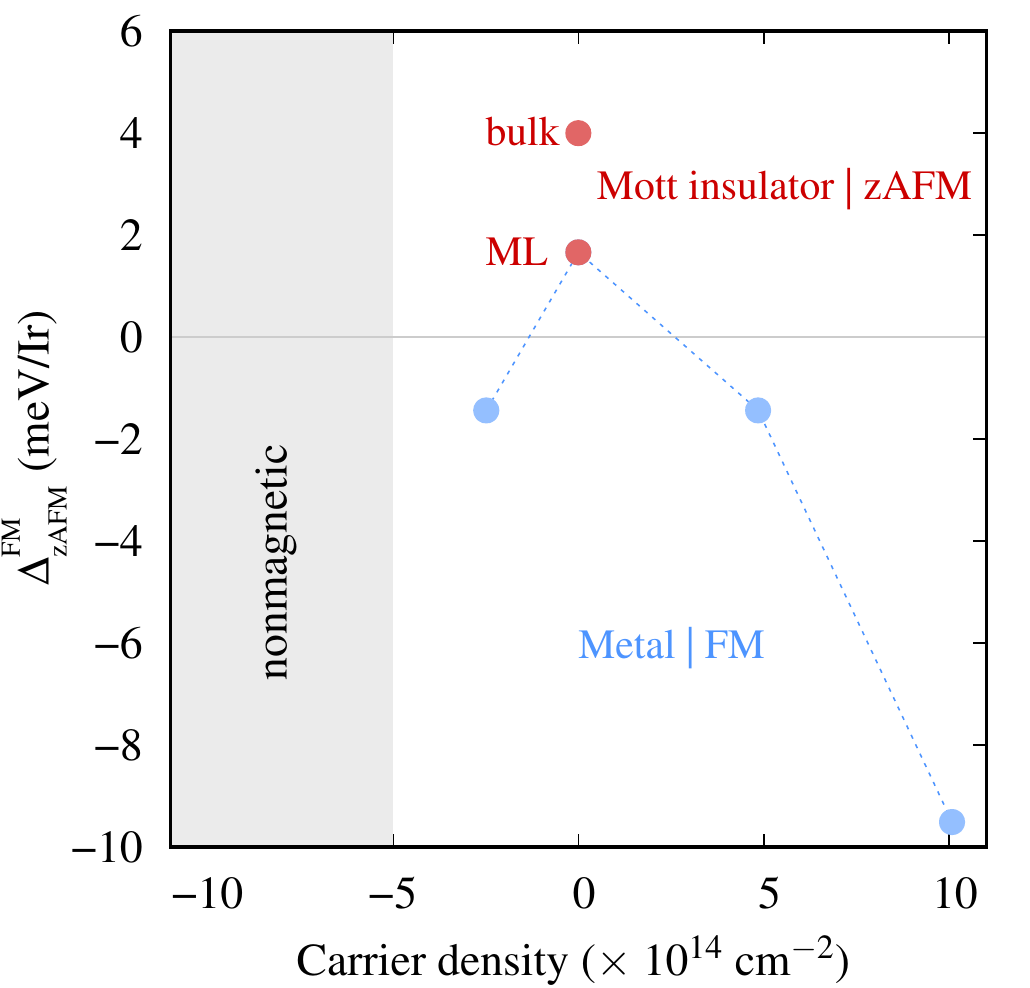}}
\caption{Relative energy of the FM phase compared to the zigzag AFM as the carrier density varies. The monolayer undergoes electronic and magnetic phase transitions under the influence of the charge carrier.  
}
\label{fig:figure6}
\end{center}
\end{figure}

We now focus on the unpassivated flakes, which can be represented in several ways based on their surface terminations. Moreover, partial passivation enables control over the charge carrier density, which induces intriguing phase transitions. Both electron and hole doping drives the Mott insulator to metal transition and triggers AFM to FM phase transition.

The charge-doped monolayer exhibits a decrease in its in-plane lattice parameters, with a relatively stronger dependence observed in the hole-doped flakes. Carrier density also affects other structural parameters such as the \ce{Ir-O} bonds, \ce{O-Ir-O} angles, and \ce{Ir-O-Ir} angles. With the addition of charge carriers, the metallic FM state becomes apparent and becomes more stable as the density increases, as depicted in Figure~\ref{fig:figure6}. However, there is no indication of magnetism in the high electron-doped [\ce{Na3|NaIr2O6|Na3}] monolayer. The moments are correlated with the carrier density. As the carrier density increases, the $\mu_{\rm o}$ decreases, and at the highest hole density, it becomes fully suppressed. This results in pure spin doublets that are coupled through Heisenberg interactions. It is worth noting that, under the influence of the charge carrier, $\mu_{\rm s}$ aligns with the $c$-direction and is collinear with $\mu_{\rm o}$. This behavior is distinct from that observed in the bulk and neutral monolayer.

\begin{figure}[!t]
\begin{center}
{\includegraphics[width=0.49\textwidth, angle=0]{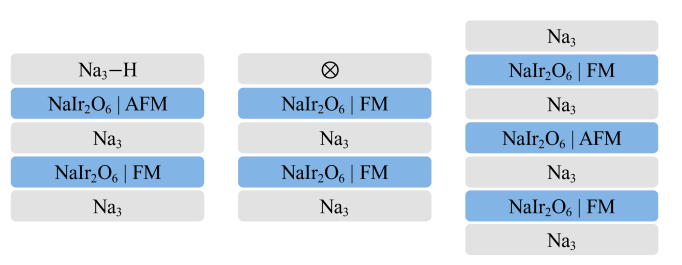}}
\caption{Magnetic ordering in the \ce{NaIr2O6} sublayers is determined by the adjacent termination layer, \ce{Na3}, passivated \ce{Na3-H} or the absence of \ce{Na3}-layer denoted by $\otimes$. These termination layers correspond to electron-doped, neutral and hole doped active layers, respectively. Schematic representation of a few, while we studied all possible terminations. The bilayer [\ce{H-Na3}$|$$\cdots$$|$\ce{Na3}] exhibits a AFM$|$FM order, while the bilayer [$\otimes$$|$$\cdots$$|$\ce{Na3}] has FM$|$FM order. The trilayer [\ce{Na3}$|$$\cdots$$|$\ce{Na3}] displays a FM$|$AFM$|$FM configuration.
}
\label{fig:figure7}
\end{center}
\end{figure}

When exposed to carrier doping, the bilayer and trilayer flakes of \ce{Na2IrO3} experience similar structural relaxations as the monolayer. The decrease in the orbital and spin moments also occurs with increased carrier density. However, the magnetism in the bilayer and trilayer is complex, and we observe the simultaneous presence of Mott insulating AFM and metallic FM order in different sublayers based on carrier density. If the magnetic sublayer of \ce{NaIr2O6} is exposed to extra electrons or holes, it transitions to a metallic FM state. However, the charge-neutral magnetic sublayer remains an AFM Mott insulator.  For instance, while the passivated bilayer [\ce{H-Na3}$|$$\cdots$$|$\ce{Na3-H}] is a Mott insulator, a metallic FM state emerges when both sides of the bilayer are terminated with or without the \ce{Na3}-layers, representing electron and hole doping, respectively. 
The interlayer coupling is always AFM, while a complete FM state lies 
higher in energy. In this state, both orbital and spin moments are aligned along the $c$-direction. Similarly, for bilayers with a passivated \ce{Na3-H} layer on one side and carrier doping on the other (Figure~\ref{fig:figure7}), the system stabilizes in an AFM$|$FM phase. Trilayers with the termination of with and without \ce{Na3}-layers on both sides, representing electron-doped [\ce{Na3}$|$$\cdots$$|$\ce{Na3}] and hole-doped [$\otimes$$|$$\cdots$$|$$\otimes$], give rise to a [FM$|$AFM$|$FM] configuration. 

Such multilayer flakes that are naturally doped could be employed in various spin-based devices like magnetic tunnel junctions, magnetic sensors, spin filters and valves, and memory devices. Nevertheless, it is important to note that devices containing \ce{Na2IrO3} flakes will only function at cryogenic temperatures. Thus, for practical application purposes, future research should prioritize finding new materials to produce ultrathin flakes capable of magnetic ordering at relatively higher temperatures.

\subsection{Minimal Heisenberg-Kitaev model}
Significant lattice distortion away from the ideal cubic octahedral environment renders conventional Heisenberg exchange, and a pure Kitaev Hamiltonian is inadequate to describe the physics. Instead, a Heisenberg-Kitaev (HK) model is appropriate for describing magnetism in spin-liquid candidate materials.~\citep{PTPS.160.155,PhysRevLett.102.017205,PhysRevLett.105.027204,PhysRevLett.110.097204,PhysRevLett.112.077204} Many variants of the HK model have been assumed in the literature; we consider a minimal HK Hamiltonian of the form,  
\begin{equation}
\begin{split}
\altmathcal{H} = \sum_{\langle ij \rangle \in \alpha \beta (\gamma)} \Bigl[  J_1 \bm{S}_i\cdot \bm{S}_j + KS_i^{\gamma}S_j^{\gamma}  & + \Gamma \left(S_i^{\alpha}S_j^{\beta} + S_i^{\beta}S_j^{\alpha}\right)\Bigr] \\
& + \sum_{\langle \langle \langle ij \rangle \rangle \rangle} J_3 \bm{S}_i\cdot \bm{S}_j, \nonumber
\end{split}
\end{equation}
where $J_1$ and $J_3$ are the first and third neighbour Heisenberg interactions, $K$ is Kitaev exchange and $\Gamma$ is the symmetric off-diagonal exchange. $\bm{S}$ are the \ce{Ir^4+} Kramers doublet pseudospin ($J_{\rm eff} = 1/2$). Other symmetric off-diagonal exchange can also be considered and has the form, $\Gamma'\left(S_i^{\beta}S_j^{\gamma} + S_i^{\gamma}S_j^{\beta} +  S_i^{\gamma}S_j^{\alpha} + S_i^{\alpha}S_j^{\gamma} \right)$. $\{\alpha, \beta, \gamma \} = \{x, y, z\}$ are three nearest-neighbour bonds. The non-Kitaev terms in the Hamiltonian trigger a phase transition at 15 K to a zigzag antiferromagnetic state.~\citep{PhysRevB.82.064412,PhysRevB.85.180403,PhysRevB.83.220403} Stronger $J_1, J_3, \Gamma$ and $\Gamma'$ will drive the system further away from the quantum spin liquid state. Therefore, exploring how the exchange interactions evolve from the bulk \ce{Na2IrO3} to a single layer of spin-orbit coupled correlated electrons is interesting.

\begin{table}[!t]
\caption{Heisenberg exchange couplings ($J_1$ and $J_3$), off-diagonal exchanges ($\Gamma$ and $\Gamma'$), and Kitaev interaction in meV; calculated for the bulk \ce{Na2IrO3} and passivated monolayer [\ce{H-Na3|NaIr2O6|Na3-H}].  The calculated N\'eel temperature $T_{\rm N}$ for the bulk exhibits excellent agreement with experimental data,~\citep{PhysRevB.82.064412,PhysRevB.85.180403,PhysRevB.83.220403} confirming the accuracy of the calculations. Surprisingly, the magnetism is strengthened in the monolayer.}
\begin{tabular}{L{0.9cm} C{1.1cm} C{1.1cm} C{1.1cm} C{1.1cm} C{1.0cm} C{1.2cm}}
\toprule
\toprule
\vspace{0.1cm}
          &  $J_1$   &  $K$       &   $\Gamma$    &   $\Gamma'$ &  $J_3$ & $T_{\rm N}$ (K)   \\
\midrule                
Bulk      &  7.22    &  $-$31.26  &   3.52        &   $-$3.24   &  3.56 & 18     \\
ML        &  6.96    &  $-$29.50  &   10.00       &   $-$5.96   &  6.14 & 32     \\
\bottomrule
\hline
\end{tabular}
\label{table1}
\end{table}

We calculate the minimal HK Hamiltonian parameters within the four-state method (Table~\ref{table1}).~\cite{PhysRevB.84.224429,PhysRevB.98.094401} The Kitaev interaction for the bulk \ce{Na2IrO3} is ferromagnetic and is the dominant energy scale, while the nearest-neighbour $J_1$  is antiferromagnetic and significantly weaker, $|J_1| \ll K$ (Table~\ref{table1}). These results are consistent with {\em ab initio} calculations~\citep{PhysRevLett.112.077204,PhysRevB.98.094401,PhysRevB.93.214431,PhysRevB.96.064430} and the exchange interactions extracted from the experimental data.~\citep{PhysRevLett.110.097204,Katukuri_2014} Further, anisotropy in the magnetic susceptibility implies the existence of symmetric off-diagonal exchange,~\citep{PhysRevB.82.064412,PhysRevLett.112.077204} and fitting with the experimental data indicates non-negligible $\Gamma$ at the limit $K \gg |J_1|$. We have estimated the ratio of $\Gamma/(3J_1 + K)$ to be approximately $-$0.37 in our current calculations for bulk \ce{Na2IrO3}, which agrees well with the previously predicted value of $-$0.3.~\citep{PhysRevLett.112.077204} It has been contended that the zigzag AFM phase in honeycomb iridates is stabilized by the third-neighbour Heisenberg exchange $J_3$, which is attributed to the presence of an inversion centre and finite \ce{M-L-L-M} hopping in the system.~\citep{Katukuri_2014,Winter_2017,PhysRevB.96.064430} Similar to $J_1$, we find $J_3$ to be antiferromagnetic and sizeable, $J_3 \sim \Gamma$ and $J_3/J_1 \sim 0.5$. Additionally, we computed the off-diagonal exchange $\Gamma'$, which is ferromagnetic unlike $\Gamma$, but comparable in strength, $|\Gamma'| \sim |\Gamma|$. Therefore, despite having a significant Kitaev interaction, the bulk \ce{Na2IrO3}  orders in the zigzag AFM phase due to the sizeable AFM Heisenberg exchanges $J_1$, $J_3$ and off-diagonal exchanges $\Gamma$ and $\Gamma'$ (Table~\ref{table1}). Corresponding $T_{\rm N}$ is calculated using the Heisenberg-Kitaev Monte Carlo using $10^4$ spins and parameters from the Table~\ref{table1}, which is in excellent agreement with the experimental data.~\citep{PhysRevB.82.064412,PhysRevB.85.180403,PhysRevB.83.220403}

Exploring the interaction parameters within the HK Hamiltonian at the limit of a monolayer is intriguing. Compared to the bulk, the electronically equivalent passivated monolayer is a spin-orbit coupled Mott insulator, albeit with specific structural alterations. Proceeding from the bulk to monolayer, the character of the interactions remains unaltered, FM $K$ and $\Gamma'$; while $J_1$, $J_3$ and $\Gamma$ are AFM (Table~\ref{table1}). Although Kitaev $K$ and Heisenberg $J_1$ remain unchanged, the magnitudes of the symmetric off-diagonal interactions $\Gamma, \Gamma'$ and third neighbour $J_3$ significantly increase in the passivated monolayer, such that $|K/J_1| \sim 4.2$, $J_3/J_1 \sim 0.9$, and $|\Gamma| > |J_3| \sim |\Gamma'|$. Compared to the bulk, the direct metal-metal hoping decreases in the monolayer while the trigonal distortion increases, and thus the ligand-assisted exchange dominates. Therefore, $\Gamma$ and $\Gamma'$ become stronger in the monolayer due to larger \ce{Ir-O-Ir} and \ce{O-Ir-O} angles.   As a result, the monolayer exhibits a pronounced stabilization of the zigzag AFM phase, evidenced by a significant increase in $T_{\rm N}$ (Table~\ref{table1}). This result contradicts the conventional wisdom that magnetism becomes weaker in the two-dimensional limit,~\citep{s41565-019-0438-6} due to the reasons discussed.

\section{Summary}

We study magnetism in ultrathin \ce{Na2IrO3} flakes in the context of the bulk being a proximate quantum spin-liquid. The magnetism persists up to the monolayer limit, and the stronger Heisenberg and off-diagonal exchange interactions further reinforce the zigzag AFM magnetism. Passivated flakes remain an AFM Mott insulator,  but an insulator-to-metal transition is triggered in naturally carrier-doped flakes, which is also accompanied by the AFM to FM transition. Hence, the multilayer flakes have the potential to be used as spin devices operating at low temperatures. The current findings provide opportunities for comprehending magnetism in two dimensions of non-van der Waals correlated oxides. Moreover, they emphasize the need for experimental assessment of magnetism and device applications.

\section{Computational Details}
First-principles calculations are performed within the density functional theory as implemented in the Vienna {\em ab initio} simulations package.~\citep{PhysRevB.47.558,PhysRevB.54.11169} The projector augmented wave formalism describes wave functions, with a plane-wave basis and a kinetic energy cut-off of 475 eV.~\citep{PhysRevB.50.17953} The Perdew-Burke-Ernzerhof functional is utilized to express the exchange-correlation energy.~\citep{PhysRevLett.77.3865} Additionally, on-site Coulomb ($U_{\ce{Ir}} = 1$ eV) and exchange ($J_{\rm H}$ = 0.5 eV) interactions for the \ce{Ir}-$d$ electrons are incorporated.~\citep{PhysRevB.52.R5467} The first Brillouin zone is sampled using a $\Gamma$-centred Monkhorst-Pack $k$-grid of 7$\times$4$\times$7 for the bulk, and 7$\times$4$\times$1 for the two-dimensional flakes.~\citep{PhysRevB.13.5188} Complete optimization of the structures is achieved by ensuring that all force components are below 5$\times$10$^{-3}$ eV/\AA, and that the total energy convergence threshold is 10$^{-8}$ eV. Charge neutral flakes are designed by passivating the surface of \ce{Na3}-layers with H. A perpendicular vacuum space of at least 20 \AA\ is employed to ensure negligible interactions between periodic images of the two-dimensional flakes. We compute the magnetic interaction parameters using the modified four-state energy method,~\citep{PhysRevB.84.224429} and to avoid spurious interactions between the ordered and manipulated spins, we employ a 2$\times$2$\times$1 supercell. To assess the dynamical stability, phonon band structures are calculated using the finite difference method within the supercell approach, and the force constants are obtained using the Phonopy code.~\citep{TOGO20151} The phonon calculations utilize a higher kinetic energy cut-off of 700 eV. N\'eel temperature is calculated using the Heisenberg-Kitaev Monte Carlo using $10^4$ spins within the SpinW code.~\citep{Toth_2015}

\section{Acknowledgements} 
M. K. acknowledges Dr Ashna Bajpai for their valuable discussions that began with the first-ever exfoliation of \ce{Na2IrO3} flakes in her group, ultimately leading to the commencement of this theoretical investigation. We gratefully acknowledge the support and resources provided by the PARAM Brahma Facility at the Indian Institute of Science Education and Research, Pune, under the National Supercomputing Mission of the Government of India. We also acknowledge funding from the National Mission on Interdisciplinary Cyber-Physical Systems (NM-ICPS) of the Department of Science and Technology, Government of India, through the I-HUB Quantum Technology Foundation, Pune, India. DKR acknowledges CSIR India for support in the form of a research fellowship.


%

\end{document}